\documentclass[twocolumn,prb, showpacs, superscriptaddress]{revtex4}
\usepackage{amssymb}
\usepackage{amsmath}
\usepackage{bbm}
\usepackage{graphicx}
\usepackage{color}

\begin{document}

\newcommand{\kh}[1]{\textit{#1}}

\author{Philipp Werner}
\affiliation{Department of Physics, University of Fribourg, 1700 Fribourg, Switzerland}
\author{Karsten Held}
\affiliation{Institut for Solid State Physics, Vienna University of Technology, 1040 Vienna, Austria}
\author{Martin Eckstein}
\affiliation{Max Planck Research Department for Structural Dynamics, University of Hamburg-CFEL, Hamburg, Germany}
\title{Role of impact ionization in the thermalization of photo-excited Mott insulators}

\date{\today}

\hyphenation{}

\begin{abstract}
We study the influence of the pulse energy and fluence on the thermalization of photo-doped Mott insulators. If
the Mott gap is smaller than the width of the Hubbard bands, the kinetic energy of individual carriers can be large enough to produce additional doublon-hole pairs via a process analogous to impact ionization. The thermalization dynamics, which involves an adjustment of the doublon and hole densities, thus changes as a function of the energy of the photo-doped carriers and exhibits two timescales: a fast relaxation related to impact ionization of high energy carriers  and a slower timescale associated with higher-order scattering processes. The slow dynamics depends more strongly on the gap size and the photo-doping concentration.   
\end{abstract}

\pacs{71.10.Fd}

\maketitle

\section{Introduction}

The photodoping of a Mott insulator provides a relatively simple way to induce and study a nonequilibrium phase transition. If a laser pulse with a frequency larger than the Mott gap is applied, doublon-hole pairs are 
produced, and  
these mobile carriers lead to a metallic response of the photodoped  Mott insulator.\cite{Iwai2003,Okamoto2007,Okamoto2010} The changes in the optical conductivity associated with this metallization have been studied experimentally using time-resolved spectroscopy. In the pioneering work by Iwai and collaborators on a Ni-chain compound,\cite{Iwai2003} a Drude peak in the conductivity was measured promptly after the photodoping pulse, and the metallic state was found to last for a few picoseconds. Alternatively, photoemission spectroscopy can be employed as a probe of the metallized Mott insulator, as was shown for 1T-TaS$_2$. \cite{Perfetti2006,Petersen2011}

One can  
distinguish two mechanisms which play a role in the relaxation of photo-doped carriers: On the one hand, electron-electron scattering can lead to a thermalization of the electronic subsystem at a hot ``electron temperature'', and on the other hand, carriers can dissipate their initially high kinetic energy through scattering with ``external"  degrees of freedom  such as spins or phonons.
A large body of theoretical work on photodoped Mott insulators has focused on the latter relaxation processes involving the scattering with spins in an antiferromagnetic background\cite{Kogoj2014,Golez2014, Eckstein2014, Eckstein2014b} or the coupling to 
phonons.\cite{Golez2012,Matsueda2012,Werner2014} 
In this paper we assume that electron-electron scattering is the fast mechanism, so that we can study the thermalization of  isolated electrons and neglect the aforementioned energy loss processes which affect the dynamics only on longer times; limits of this assumption will be discussed in more detail below. Generally, this assumption is valid if the electron-phonon coupling strength is weak and if there are no  spin correlations, e.g., due to a high temperature 
or large fluence.

In metals, a rapid thermalization of the electronic system is typically observed and underlies the assumption that a quasi-equilibrium 
picture or two temperature model\cite{Allen1987} 
can be used for describing the dynamics already at very short times after an excitation. 
In an insulator, the thermalization and relaxation involves an adjustment in the number of electron-hole pairs, 
which can be a slow process in the presence of a large gap. It was found that in a purely electronic model 
(a paramagnetic Hubbard  
model with on-site repulsion $U$) the thermalization time 
depends exponentially on the gap size, and that even the relaxation of the distribution of photodoped carriers within the Hubbard bands can be extremely slow.\cite{Eckstein2011pump} The explanation for this is relatively simple. 
If the energy $U$ which is needed for the production of a single doublon-hole pair is substantially larger than the typical kinetic energy of a single doublon or hole, complicated multi-particle scattering processes are needed for thermalization -- hence the exponential scaling with $U$.\cite{Sensarma2010} 
For a similar reason, the doublon-hole recombination via emission of magnons \cite{Lenarcic2013} or phonons \cite{Mitrano2014} becomes slow when the gap is large.

If the kinetic energy of the charge carrier (doublon or hole) is larger 
than the size of the Mott gap, it is energetically allowed to create
an {\em additional}  doublon-hole pair via  two-particle scattering.
We will call such processes ``impact ionization",
in analogy to similar processes in semiconductors\cite{SCimpact} 
and atoms.\cite{ATOMimpact}
Since a large excess kinetic energy is needed, one may anticipate a strong dependence 
of the relaxation dynamics on the pulse energy (for a fixed interaction or gap size). In particular, we may encounter a situation where photo-doped doublons inserted at the upper edge of the upper Hubbard band trigger a rapid increase in the number of charge carriers through impact ionization, while for doublons inserted at the lower band edge, the kinetic energy is not sufficient for impact ionization, so that the doublon-hole production  depends on rare multi-particle scattering events.

Since impact ionization processes have the potential to rapidly enhance the  number of mobile carriers in the photo-induced metal,  
an understanding of this physics is crucial for possible applications of photo-induced metal-insulator transitions in ultra-fast switches, or for the efficient operation of photovoltaic devices. For example, impact ionization allows to create multiple doublon-hole pairs per photon and hence to overcome 
the Schockley-Queisser limit
for the efficiency of solar cells.\cite{SchockleyQueisser61,Manousakis2010}

In this paper 
we focus on the paramagnetic 
Mott-Hubbard  insulator  
with a relatively small gap, and study in more detail the electronic thermalization processes, with the goal of disentangling the fast impact ionization channel from doublon-doublon scattering and  slower multi-particle scattering processes. 
As we will show, 
impact ionization can be identified experimentally by characteristic signatures in the time-resolved photo-emission spectrum, and in particular its energy and fluence dependence: 
(i) Impact ionization can increase the number of doublons on the 10 fs time scale, and thus result in a rapid spectral weight increase of up to a factor of three above the Fermi level. 
(ii)  While the photoemission spectral weight at high energies 
decreases as a function of time, 
it increases more than proportionally at energies which are lower by at least the size of the gap. (iii) The
frequency of the pump pulse needs to be larger than twice the gap; 
below this threshold impact ionization is not possible.
(iv) Since impact 
ionization involves only a single doublon or hole this process
does not depend strongly  on the density of photo-doped carriers (or the fluence), while doublon-doublon and higher order scattering processes will become more frequent if the density of photo-induced carriers increases.

\section{Model and method}

We investigate and quantify the effect of impact ionization by considering a Hubbard model 
\begin{equation}
H=\sum_{ij,\sigma}v_{ij}c^\dagger_{i\sigma}c_{j\sigma}+U\sum_i(n_{i\uparrow}-\tfrac12)(n_{i\downarrow}-\tfrac12)
\end{equation}
with on-site interaction $U$ comparable to the bandwidth. The operators $c_{i\sigma}$ create an electron at site $i$ with spin $\sigma$, and the hopping amplitude is $v_{ij}$. The model is solved on an infinite-dimensional hypercubic lattice using nonequilibrium dynamical mean field theory (DMFT)\cite{Freericks2006, Aoki2013} with a strong-coupling perturbative impurity solver (non-crossing approximation, NCA).\cite{Eckstein2010nca} This lattice has a Gaussian density of states, $\rho(\epsilon)=1/(\sqrt{\pi}W)\exp(-\epsilon^2/W^2)$, and we use the width $W$ as our unit of energy. To simulate the photo-doping pulse, we apply a few-cycle electric field pulse of the form
\begin{equation}
E(t)=E_0e^{-(t-t_p)^2/\sigma^2}\sin(\Omega(t-t_p))
\end{equation}
with $t_p=6$ and $\sigma^2=6$ in the body-diagonal of the lattice. We use a gauge without scalar potential, so that the field is given by the time derivative of the vector potential $A$, $E(t)=-\partial_t A(t)$. Using the Peierls substitution, the field then enters the Hamiltonian via a time-dependent shift of the dispersion, $\epsilon_k\rightarrow \epsilon_{k-A(t)}$, where $\epsilon_k$ is the Fourier transform of the hopping matrix. For details of our implementation of the nonequilibrium DMFT equations, the treatment of the electric field and the NCA impurity solver, we refer to 
Refs.~\onlinecite{Eckstein2011pump} and \onlinecite{Aoki2013}. The double occupation at time $t$, $d(t)$, is a local observable which can be obtained directly from the solution of the effective impurity model. Since we use a strong-coupling impurity solver, $d(t)=\mathcal{G}^<_{|\uparrow\downarrow\rangle}(t,t)$ is simply the occupation of the pseudo-particle state corresponding to doubly occupied sites.\cite{Eckstein2010nca}  
Because $d$ is nonzero already in the initial Mott state due to virtual charge fluctuations, the number of photo-excited doublons at a later time $t$ is given by the difference $d(t)-d(0)$. This number can also be obtained spectroscopically, by integrating the photoemission spectrum over the upper Hubbard band (see below).

The relaxation dynamics of photo-doped carriers depends crucially on whether or not the Mott insulator is antiferromagnetically 
ordered.\cite{ Werner2012, Eckstein2014, Eckstein2014b} 
In the present study, we restrict the calculations to paramagnetic  photodoped Mott insulators
at elevated temperatures. 
The temperature considered in this paper, i.e, $T=1/\beta=1/5$, is well above the highest DMFT N\'eel temperature, $T_N=1/7$.\cite{Jarrell92a} Since DMFT overestimates $T_N$, we can be certain that also paramagnons are not important. This is also supported by recent non-equilibrium cluster DMFT calculations \cite{Eckstein2014b} which, for the 2D Hubbard model, showed that short-ranged spin-correlations have an important effect on the  
relaxation dynamics only below  $T=(1/5)\; 4t_*$.\cite{energyscales} 
(For higher-dimensional lattices, the effect of short-range correlations in the paramagnetic 
phase can be expected to be even smaller.)

\section{Results}

\subsubsection{Pulse-frequency dependence}
\label{subsec_pulse_freq}

\begin{figure}[t]
\begin{center}
\includegraphics[angle=-90, width=0.9\columnwidth]{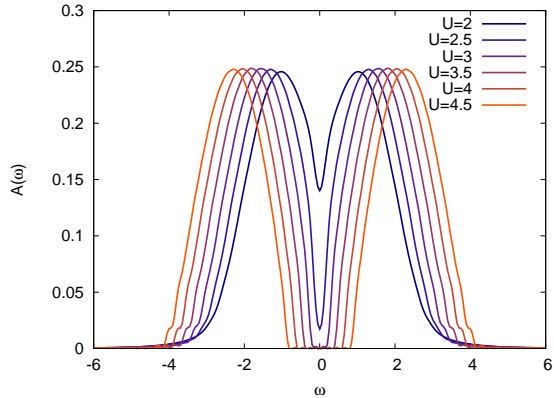} 
\caption{Equilibrium spectral functions of the Hubbard model for $\beta=5$ and indicated values of $U$ as calculated by DMFT(NCA).}
\label{equilibrium}
\end{center}
\end{figure}

For the purpose of orientation we first plot the equilibrium spectra for different values of $U$ and inverse temperature $\beta=5$ (Fig.~\ref{equilibrium}). The gap opens at $U\approx 2.5$ and then grows approximately linearly with $U$. In the insulating phase, the shape of the Hubbard bands is almost independent of $U$, and they have a width of about $3$. 
Since an impact ionization process involves the scattering of a doublon at the upper edge of the upper band to the lower edge and a simultaneous doublon-hole excitation, it already becomes clear from the equilibrium spectra that we can only expect these processes to be relevant for interactions $U\lesssim 4$. 
For larger $U$ values, the energy associated with scattering between states within the band 
is not enough to excite electrons across the gap. 
In the following we will thus focus on the interaction range $2.5\le U \le 4$.

\begin{figure}[t]
\begin{center}
\hfill
\includegraphics[angle=-90, width=0.465\columnwidth]{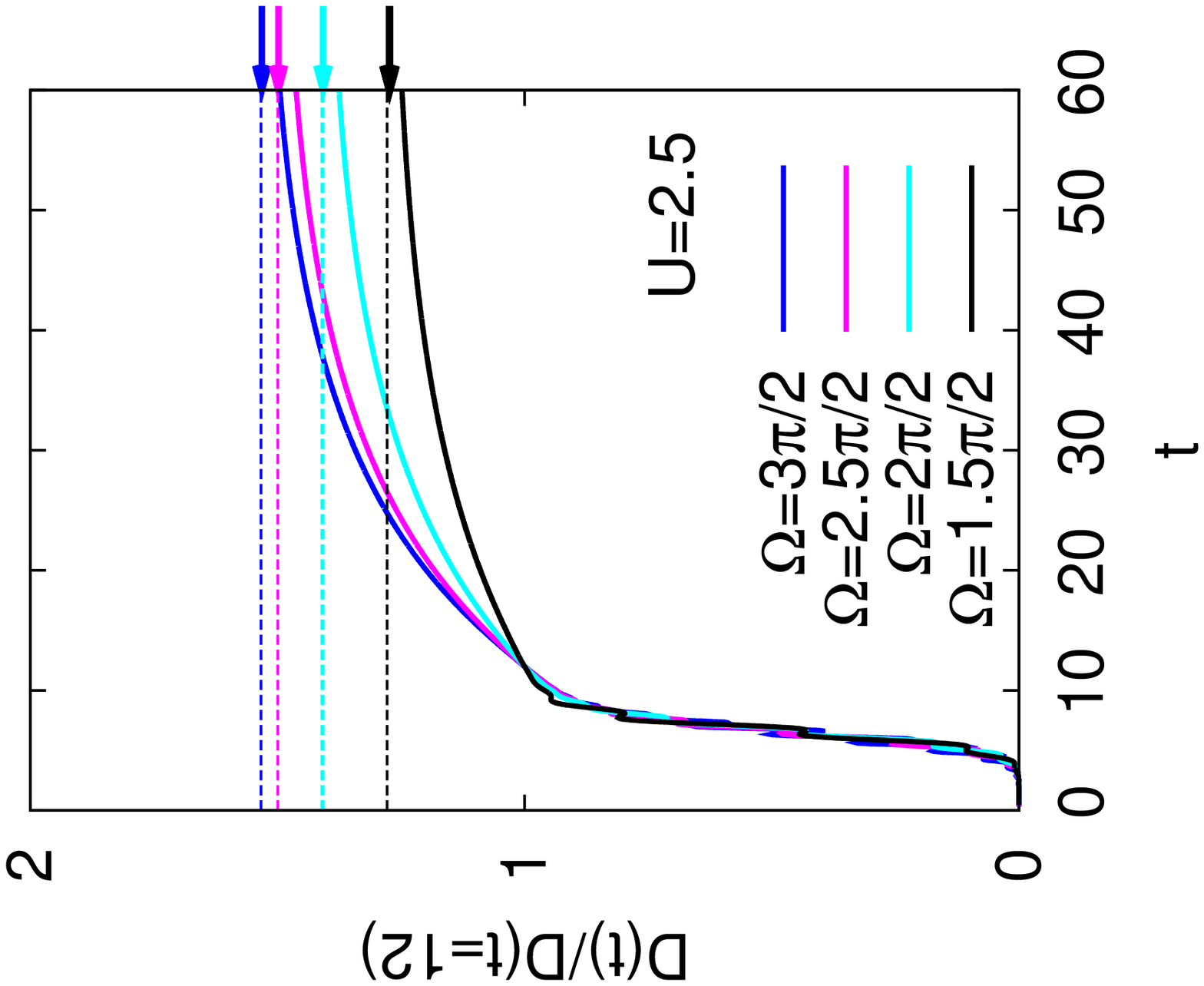}
\includegraphics[angle=-90, width=0.465\columnwidth]{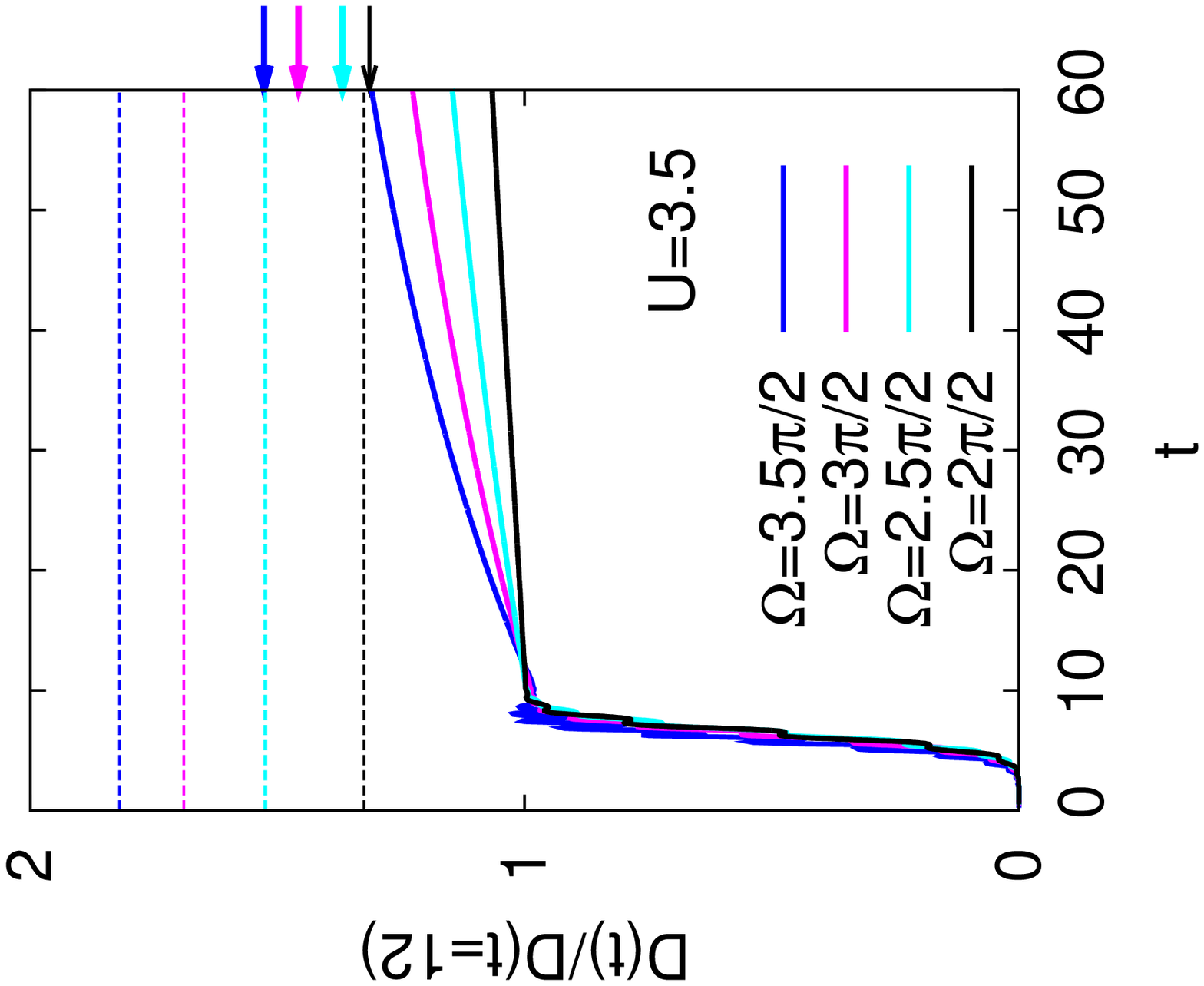}\\
\includegraphics[angle=-90, width=\columnwidth]{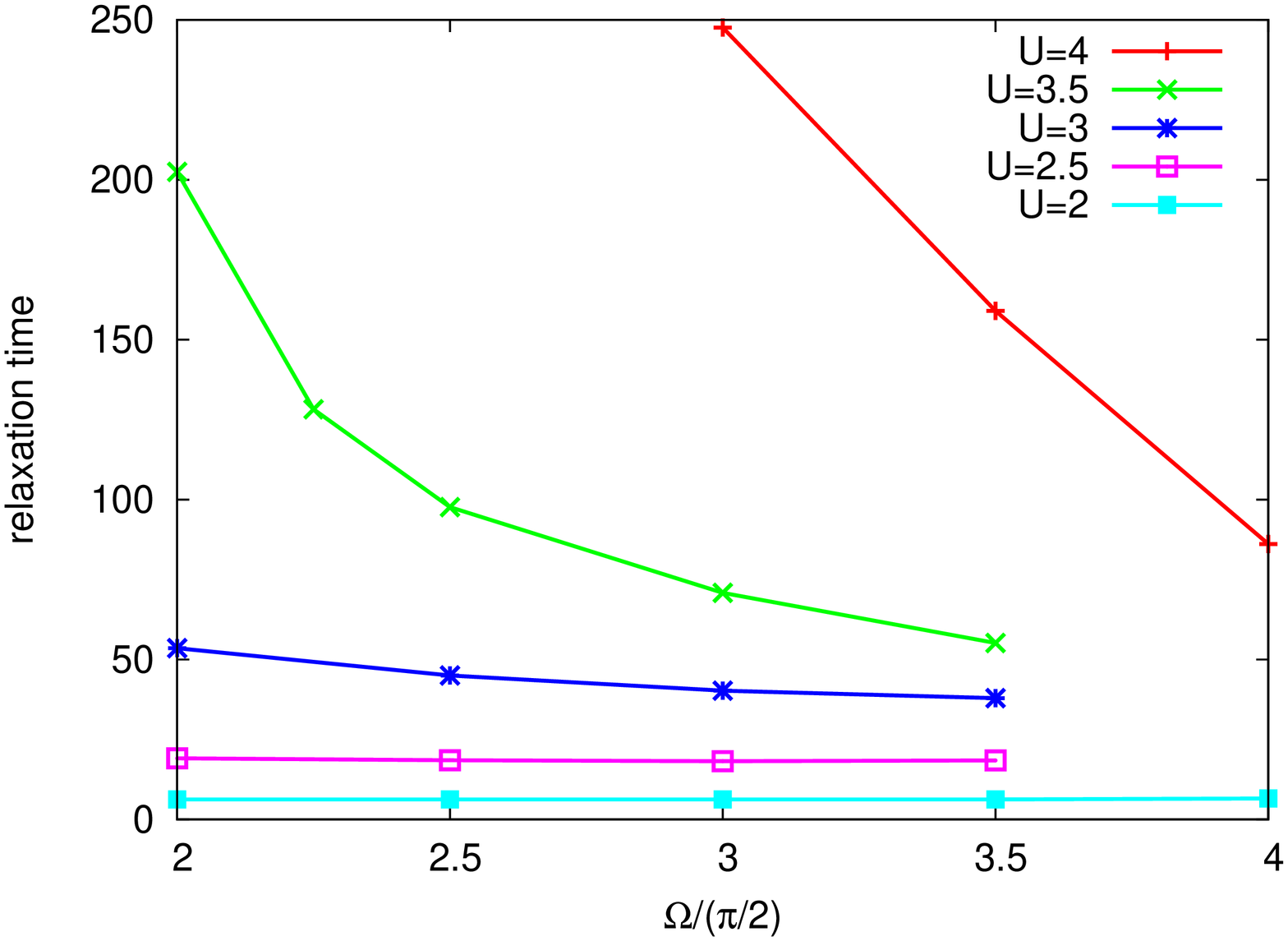}
\caption{Relaxation after pulse excitations with different frequencies at an initial inverse temperature $\beta=5$. The amplitude of the pulses is adjusted such that the number of photo-doped doublons at $t=12$ (shortly after the pulse) is $0.01$.  
Top panels: time evolution of the normalized doublon density and expected thermal values (horizontal lines) for $U=2.5$ and $3.5$. Bottom panel: relaxation times ($c$) obtained by fitting 
$D(t)/D(t=12)$ 
to the function $a+b\exp(-t/c)$ in the range $t\in[30,60]$. The extrapolated long-time values ($a$) for $U=3.5$ are indicated by arrows in the upper panels.
}
\label{doublon_rate}
\end{center}
\end{figure}

For $U=2.5$ and $3.5$, the time evolution of the 
photo-doped doublon density $D$ 
after pulses with frequencies in the range 
$1.5\pi/2 \le \Omega \le 3.5\pi/2$ 
is plotted in the top panels of Fig.~\ref{doublon_rate}. Here and in the following, we examine the change of double occupation with time, $D(t)=d(t)-d(0)$. For a better comparison between different band gaps and pulse  energies, the amplitude $E_0$ of each pulse has been adjusted such that at $t=12$, shortly after the pulse, the density of photo-doped doublons is $d(t=12)=0.01$, and we normalize the curves by this initial density.  
We see that during the thermalization process, the number of doublons increases, 
i.e., excess kinetic energy of the photo-doped carriers is transformed into interaction energy. 
The thermal reference value can be calculated by measuring the energy $E_j=\int dt\, j(t) \cdot E(t)$ injected into the system by the pulse. Here, $j=\sum_k n_{k\sigma}v_k$ is the current, with 
$n_{k\sigma}(t)=-iG^<_{k\sigma}(t,t)$  
and $v_k(t)=\partial_k\epsilon_{k-A(t)}$. By comparing the total energy after the pulse to that of an equilibrium system, we can compute the temperature $1/\beta_{\rm eff}$ and double occupancy which the system will reach, assuming thermalization, in the long-time limit.  The thermal values of the double occupancy are indicated by the dashed horizontal lines in Fig.~\ref{doublon_rate}.

\begin{figure*}[ht!]
\begin{center}
\includegraphics[angle=-90, width=\columnwidth]{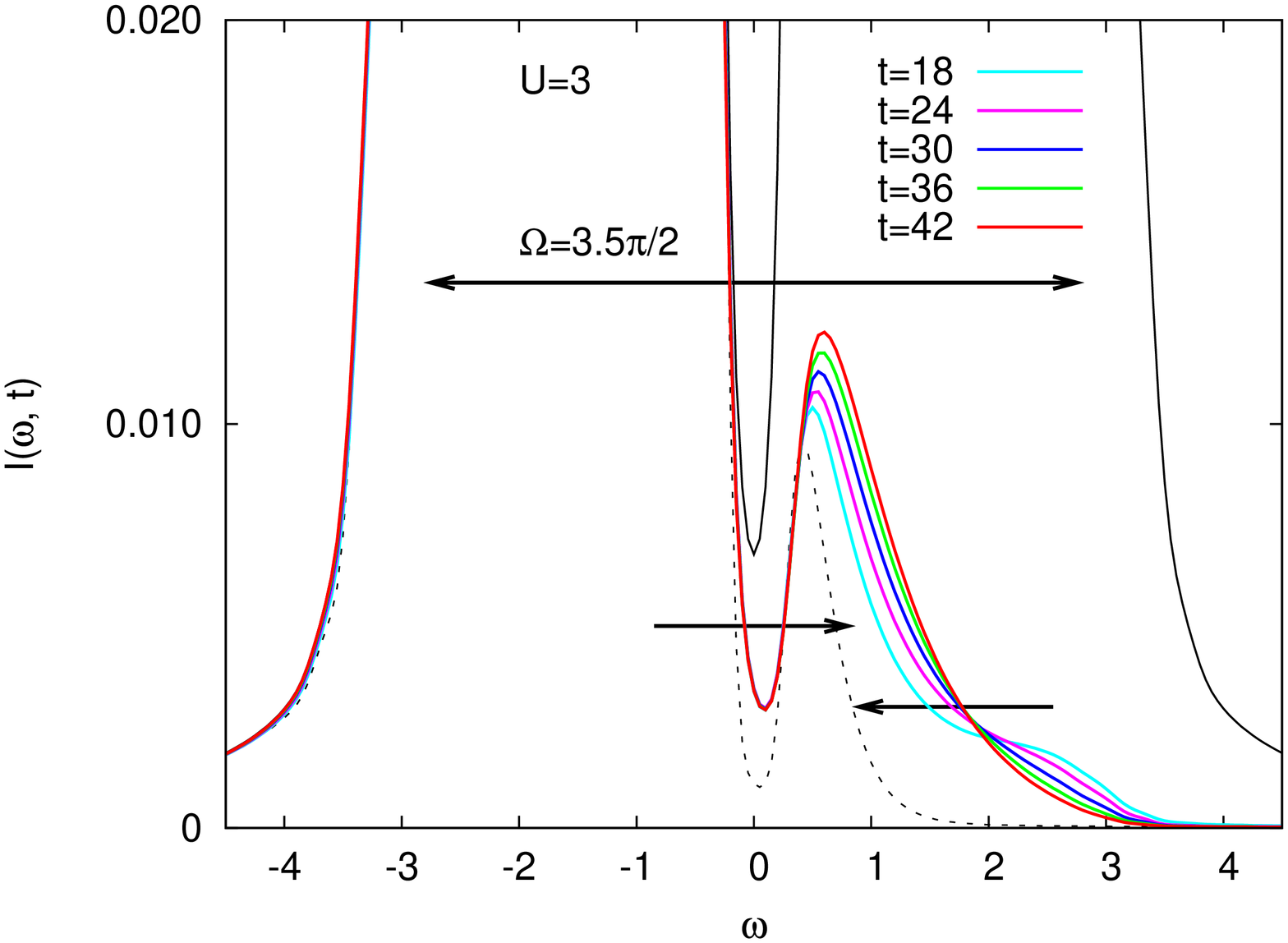}
\includegraphics[angle=-90, width=\columnwidth]{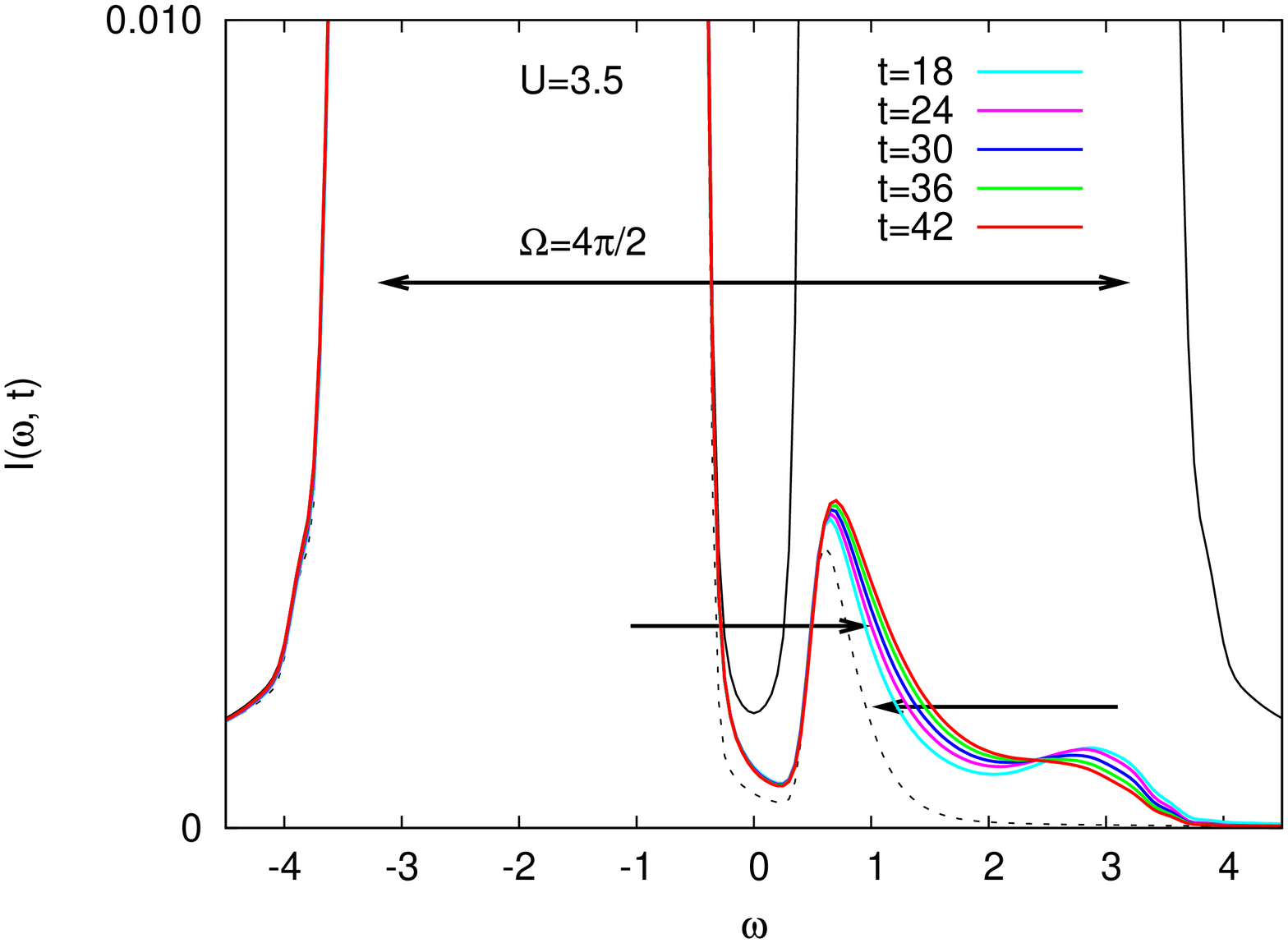}
\includegraphics[angle=-90, width=\columnwidth]{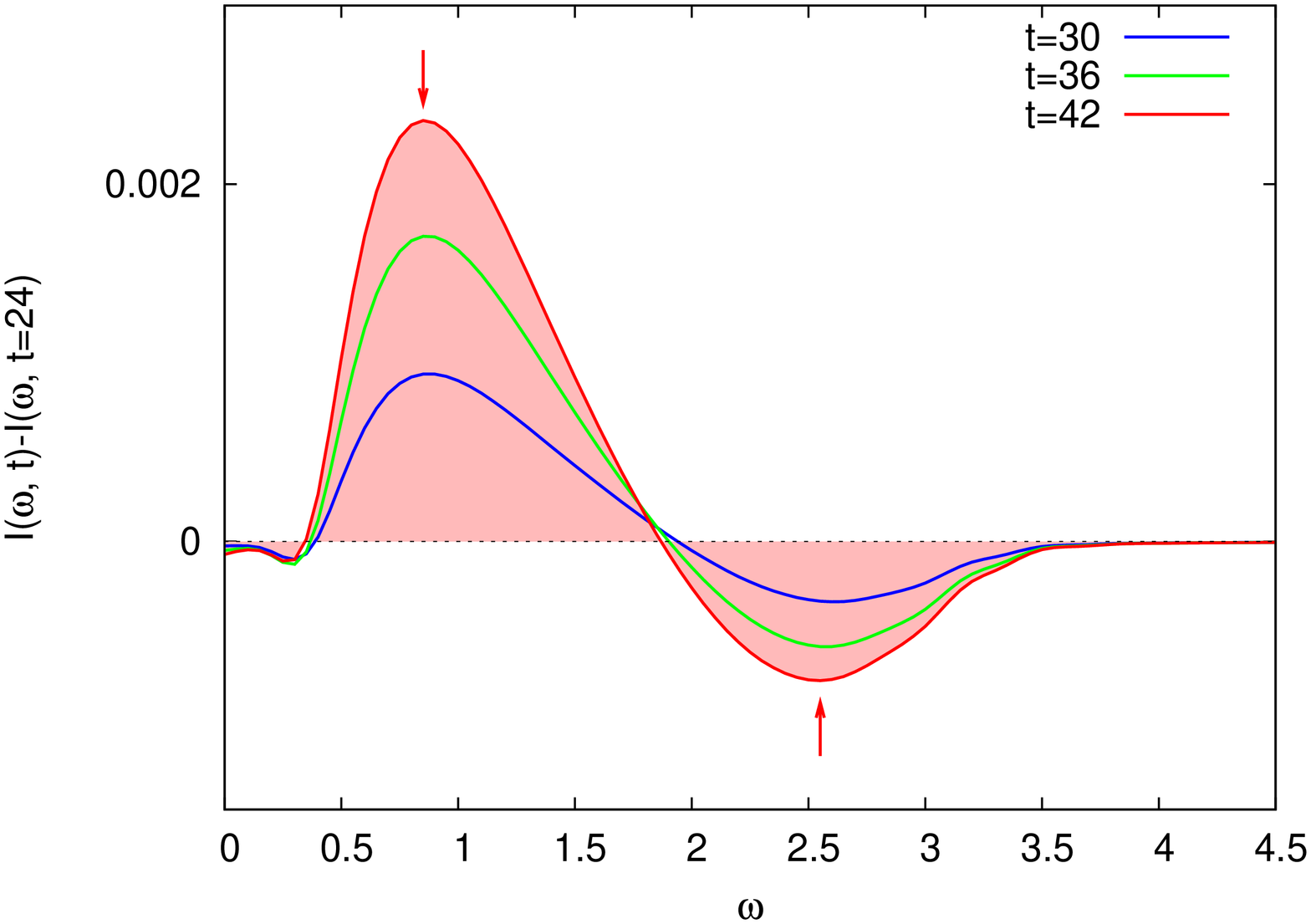}
\includegraphics[angle=-90, width=\columnwidth]{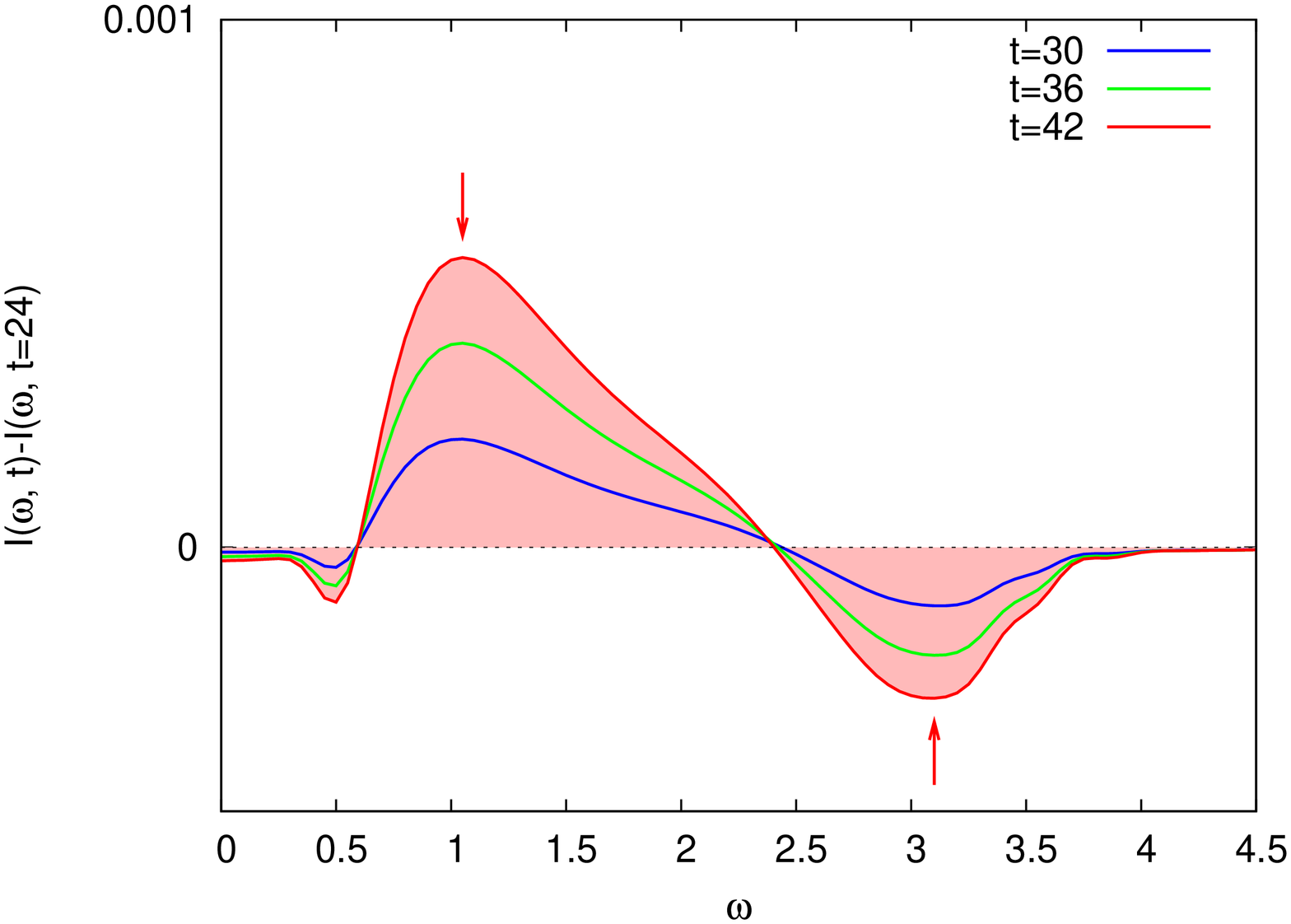} 
\caption{(Color online)
Time-resolved photoemission spectra for pulse amplitude $2$ and initial inverse temperature $\beta=5$. The left panels show results for 
$U=3$, $\Omega=3.5\pi/2$ (initial photo-doping concentration $D(t=12)=0.0056$) and the right panels for  $U=3.5$, $\Omega=4\pi/2$ (initial photo-doping concentration $D(t=12)=0.0021$).
Colored curves in the upper panels show the nonequilibrium photoemission spectrum $I(\omega,t)$ for indicated values of $t$, while the dashed black curves plot that of the initial equilibrium state. 
Solid arrows sketch the energy transfers associated with an impact ionization process: 
(left-pointing arrow: kinetic energy loss of a high energy doublon; right-pointing arrow: excitation of an electron across the gap). Lower panels: time dependent change in the photoemission spectrum. Red arrows indicate the energies $\omega_\text{gain}=0.85$ and $\omega_\text{loss}=2.55$ ($U=3$, left panel) and $\omega_\text{gain}=1.05$ and $\omega_\text{loss}=3.1$ ($U=3.5$, right panel), respectively. The red areas correspond to the increase (decrease) in low-energy (high-energy) doublons from $t=24$ to $t=42$. For $U=3$, the increase in the number of low-energy doublons is about $2.7$ times as large as the decrease in the number of high-energy doublons, while for $U=3.5$, the corresponding factor is about $2.3$.
}
\label{doubling}
\end{center}
\end{figure*}

If we fit the doublon curves in the range $30\le t\le 60$ to 
a single exponential 
$a+b\exp(-t/c)$ we obtain the relaxation times $c$ plotted in the bottom panel of Fig.~\ref{doublon_rate}, and the long-time values $a$ indicated by the arrows in the top panels. For $U\gtrsim 3$, this extrapolated thermalized double occupation  $a$  is smaller than its actual thermal value (dashed line in Fig.~\ref{doublon_rate}). Hence, we can conclude that at least  two relaxation mechanisms are at work. 
We also note that the relaxation times $c$ are much faster than
previously observed electronic thermalizations of doublons  \cite{Eckstein2011pump,Sensarma2010} and 
strongly pulse-energy dependent:
with increasing pulse frequency,  the initial growth of the doublon population becomes faster, see upper right and lower panel of Fig.~\ref{doublon_rate}.
All of this suggests that the fast doublon production is due to
impact ionization, which
requires that the excess kinetic energy of the photo-doped carriers is larger than the gap. Once all carriers with large kinetic energy have decayed, this contribution disappears and the long-time thermalization dynamics is controlled by slower multi-particle scattering processes. In previous studies, this was the only relaxation mechanism, 
since the pulse energy was too small (relative to the gap) for impact ionization.

In order to show direct evidence for impact ionization, we plot in Fig.~\ref{doubling}  time resolved photoemission spectra
\cite{FreericksTRPES} 
\begin{equation} 
I(\omega,t)=-i\int dt_1 dt_2 S(t_1)S(t_2)e^{i\omega(t_1-t_2)} G^<(t+t_1,t+t_2), 
\end{equation}
for a Gaussian probe pulse envelope $S(t)=\exp(-t^2/2\delta^2)\Theta(1.5\delta-|t|)$, which we cut off for $|t|>1.5\delta$. Choosing $\delta=12$, and a pump-pulse lasting up to $t=12$, this means that for $t>30$, there is no overlap between pump and probe pulse anymore. A pulse width of $\delta$ allows to measure the relaxation of the system with an energy resolution $\sim 1/\delta$. 

Let us first focus on the left panels, which show results for interaction $U=3$, pulse amplitude $2$ and pump pulse frequency $\Omega=3.5\pi/2$. This pump pulse inserts the doublons at the upper edge of the upper Hubbard band. As time increases, the spectral weight near the upper band edge decreases, while the weight near the lower band edge starts to grows. 
(Up to a constant off-set, the total weight in the upper Hubbard band reproduces $d(t)-d(0)$ to high accuracy.) 
Since the energy difference between the upper and lower band edge is larger than the gap size in this example, impact ionization processes can be expected to play a role in the initial relaxation dynamics. 

That these processes are indeed to a large extent responsible for the doublon production follows from the lower panel, which plots the difference between the photoemission spectrum at time $t$ and the measurement at time $t=24$. Spectral weight decreases with increasing time for $\omega\gtrsim 1.9$, with the fastest decrease measured at energy $\omega_\text{loss}=2.55$ (indicated by an arrow in  Fig.~\ref{doubling}). From the lower band edge up to $\omega\lesssim 1.9$, we see an increase in spectral weight, with the fastest doublon production at energy $\omega_\text{gain}=0.85$. 
Impact ionization can now be identified by analyzing the number of doublons produced per decay of a high energy doublon. 
Let us consider the process of a high energy doublon creating a doublon-hole pair (doublon$_{\rm high}$ $\rightarrow$ doublon$_{\rm low}$ +  doublon$_{\rm low}$ + hole$_{\rm low}$), and its symmetric counterpart  
(hole$_{\rm high}$ $\rightarrow$ hole$_{\rm low}$ +  hole$_{\rm low}$ + doublon$_{\rm low}$). 
The net effect is the production of {\it three} low energy doublons per decay of a high energy doublon. Therefore, if impact ionization were the only relevant process on the timescale of Fig.~\ref{doubling}, we would expect that the increase in the number of low energy doublons would be three times larger than the decrease in the number of high energy doublons. Computing the integrals over the positive and negative parts of the curves displayed in the bottom panel (red shaded areas), we find a ratio of $2.7$.  This indicates that besides the impact ionization processes, there are also doublon conserving scattering processes which contribute to the redistribution of spectral weight within the Hubbard band. We will discuss some key differences between these two relaxation channels, and how they affect the time-resolved photoemission spectra, in the section on the fluence dependence (Sec.~\ref{subsec-fluence}). 

A second observation is that the ratio in the positions of the maxima of the gain and loss peak is approximately  
$\omega_\text{gain}/\omega_\text{loss}=3$. A tempting interpretation would be to say that in an impact ionization process the shift of occupied spectral weight within the upper Hubbard band from $\omega_\text{loss}$ to $\omega_\text{gain}$ is associated with a transfer of occupation between the Hubbard bands from $-\omega_\text{gain}$ to $\omega_\text{gain}$ (scattering process indicated by the arrows in the upper panel). However, in contrast a band insulator the first moment of the occupied density of states in Mott insulators does not equal the total energy, so that the aforementioned redistribution of occupied weight would not be energy-conserving. Doublon-hole excitations lead to a reconstruction of the density of states and hence a redistribution of weight over a larger $\omega$ region. Here we will not analyze this effect in detail, but instead focus on the evolution of the spectral weight averaged over large energy regions (high and low energy doublons).

The right hand panels of Fig.~\ref{doubling} show analogous results for $U=3.5$, pulse amplitude $2$ and pulse frequency $\Omega=4\pi/2$. While the absorption is smaller in this case, the parameters are still compatible with impact ionization. Indeed, as shown in the lower panel, the change in the spectral function is fastest near the energies $\omega_\text{gain}=1.05$ and $\omega_\text{loss}=3.1$, which satisfy $\omega_\text{gain}/\omega_\text{loss}\approx 3$. The low-energy hump is however broader, and the ratio between the red areas is only $2.3$, which suggests a larger role of doublon-doublon and doublon-hole scattering processes in this case.

\subsubsection{Two-step thermalization}

At least in cases such as the set-up discussed above, where the high-energy and low-energy carriers can be relatively clearly separated, one can try to reproduce the time evolution of the doublon population with a simple model that 
describes the decay of the high-energy doublons via impact ionization with a relaxation time $\gamma$, and the higher order scattering processes with a different associated thermalization time $\tau$. 
We denote the slow processes with a subscript ``therm" and the fast ones with ``imp" and split the total doublon number $D$ 
into a high-energy and low-energy population $D_1$ and $D_2$, respectively. 
After thermalization, we assume that only low-energy doublons are present, and denote their number by $D_\text{th}$. 
The time evolution is then given by the equations
$\frac{dD_1}{dt}=
\left(\frac{dD_1}{dt}\right)_\text{imp}$ and
$\frac{dD_2}{dt}=\left(\frac{dD_2}{dt}\right)_\text{therm}+\left(\frac{dD_2}{dt}\right)_\text{imp}$, 
where we assume 
the simple rate equations 
\begin{align}
\Big(\frac{dD_1}{dt}\Big)_\text{imp}&=-\frac{1}{\gamma}D_1,\label{model1}\\
\Big(\frac{dD_2}{dt}\Big)_\text{imp}&=-3\Big(\frac{dD_1}{dt}\Big)_\text{imp},\label{model2}\\
\Big(\frac{d}{dt}D_2\Big)_\text{therm}&=\frac{1}{\tau}\Big(D_\text{th}-D_2\Big).\label{model3} 
\end{align}
The factor of three in Eq.~\eqref{model2} accounts for the production of three low energy-doublons 
per decay of a high-energy doublon (hole) in an impact ionization process, as explained above. 
The equations governing the time evolution of the two components thus read 
$\frac{dD_1}{dt}=
-\frac{1}{\gamma}D_1,$ 
$\frac{dD_2}{dt}=\frac{1}{\tau}(D_\text{th}-D_2)+\frac{3}{\gamma}D_1$, 
and  
the solution for the total doublon population for times $t>t_s$ becomes
\begin{align}
&D_\text{th}-D(t)=\frac{2\tau+\gamma}{\tau-\gamma} D_1(t_s) e^{-(t-t_s)/\gamma}\nonumber\\
&\hspace{5mm}+\Bigg(D_\text{th}-D(t_s)-\frac{2\tau+\gamma}{\tau-\gamma}D_1(t_s)\Bigg)e^{-(t-t_s)/\tau}.
\label{model}
\end{align}
Here, $t_s$ is some time after the pulse (we choose $t_s=15$ in the following analysis), $D_\text{th}$ and $D(t_s)$ are known, while $D_1(t_s)$, 
$\gamma$ and $\tau$ 
must be obtained by fitting.

\begin{table}[b]
\begin{center}
\begin{tabular}{llllllll}
$U$\hspace{5mm}\mbox{}  & $\Omega$\hspace{6mm}\mbox{}  & $D_\text{th}-D(t_s)$ & $\frac{D_1(t_s)}{D(t_s)}$\hspace{6mm}\mbox{} & $\gamma$\hspace{6mm}\mbox{} & $\tau$ \\
\hline
2.5 	&	$\frac{3\pi}{2}$ 	&	0.00448	& 0.0088 &	7.20	& 	18.8	 \\
2.5 	&	$\frac{2.5\pi}{2}$ 	& 	0.00421	& 0.0067 &	7.75	& 	19.0 	 \\
2.5 	&	$\frac{2\pi}{2}$ 	& 	0.00348	&  0.0044   &	 9.35 	& 19.6 \\
\hline
3 	&	$\frac{3.5\pi}{2}$ 	&	0.00684	&0.046&	13.4	& 	60.3 	 \\
3 	&	$\frac{3\pi}{2}$ 	&	0.00674	&0.040&	15.0	& 	61.4	\\
3 	&	$\frac{2.5\pi}{2}$ 	& 	0.00573	&0.026&	16.5	& 	64.9 	\\
\hline
3.5 	&	$\frac{3.5\pi}{2}$ 	&	0.00789	&0.15&	44.0	& 	376 	 \\
3.5 	&	$\frac{3\pi}{2}$ 	&	0.00669 	&0.083&	48.4	& 	257 	\\
\hline 
( 4 	&	$\frac{4\pi}{2}$ 	&	0.00820	&0.19  &	86.9	& 	5990 )\\

\hline
\end{tabular}
\end{center}
\caption{Relaxation times and initial excited populations $D_1(t_s)$ extracted from fits to model (\ref{model}) in the range $t\in [15,60]$ ($t_s=15$). 
The doping concentration after the pulse is $D(t_s)=0.010$ in all cases.
}
\label{table}
\end{table}

For $U=2.5$ the relaxation is 
well 
described by a single exponential. This follows already from the data in the top left panel of Fig.~\ref{doublon_rate}, which show that the extrapolated long-time values from an exponential fit in the range $t\in[30,60]$ correctly predict the thermal doublon density.  This is however a special case, since the gap is just opening at $U=2.5$. In this situation, additional doublons can be easily generated and the relaxation to the expected thermal value is fast.

For $U=3$ and $3.5$, a single exponential model
is not appropriate anymore, but  fitting of the data with
the double-exponential decay (\ref{model}) 
works rather well. We summarize the results of this analysis in Tab.~\ref{table}. 
One finds fast relaxation times  
$\gamma\sim 15$
and slow relaxation times $\tau\sim 60$ for $U=3$, and fast (slow) relaxation times of approximately 
$40$-$50$ ($250$-$350$) for $U=3.5$. The (relative) initial excited population decreases as the pulse frequency is lowered, in rough agreement with the time-resolved spectra. At the lowest pulse frequencies considered, the separation between high-energy and low-energy populations becomes blurred and our model fit becomes less meaningful.  
For $U=4$, all relaxation times become rather long, and it is difficult to obtain reliable fits. 
We find  $\gamma\approx 90$ and $\tau\approx 6000$ (with a large uncertainty). 

While one should probably not consider more than the first digit of the relaxation times and initial high-energy populations in Tab.~\ref{table}, 
our model does provide a consistent description of the doublon relaxation, and the results demonstrate that impact ionization processes play a significant role in the interaction range $3\le U\le 4$. In particular, they lead to a two-step thermalization with a fast initial doublon production and an associated transfer of spectral weight from the upper to the lower band edge, followed by a much slower thermalization of the relaxed distribution. 
The slow timescale $\tau$ grows fast with increasing $U$, which is consistent with a previous analysis based on single exponential fits.\cite{Eckstein2011pump} However, also the relaxation time $\gamma$ associated with the impact ionization increases with $U$, which indicates that these processes become less likely as the energy cost of producing a doublon-hole pair increases. Note that the excess kinetic energy of the doublon has to be larger than the Mott gap, which is increasing with $U$. On the other hand,  the kinetic energy of the photo-doped carriers is 
essentially bounded by the non-interacting bandwidth which is independent of $U$.

 \begin{figure}[b]
\begin{center}
\includegraphics[angle=-90, width=0.99\columnwidth]{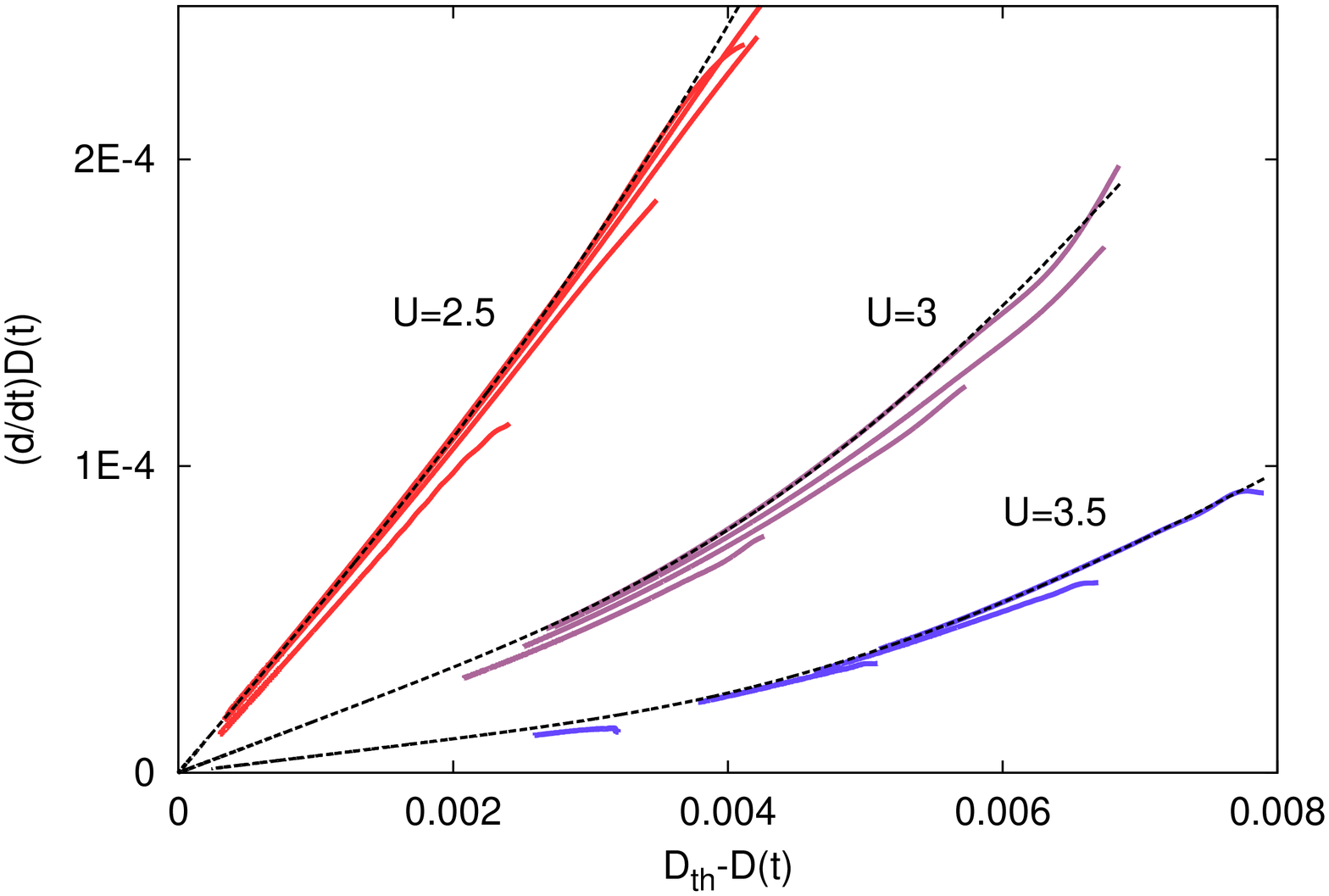}
\includegraphics[angle=-90, width=\columnwidth]{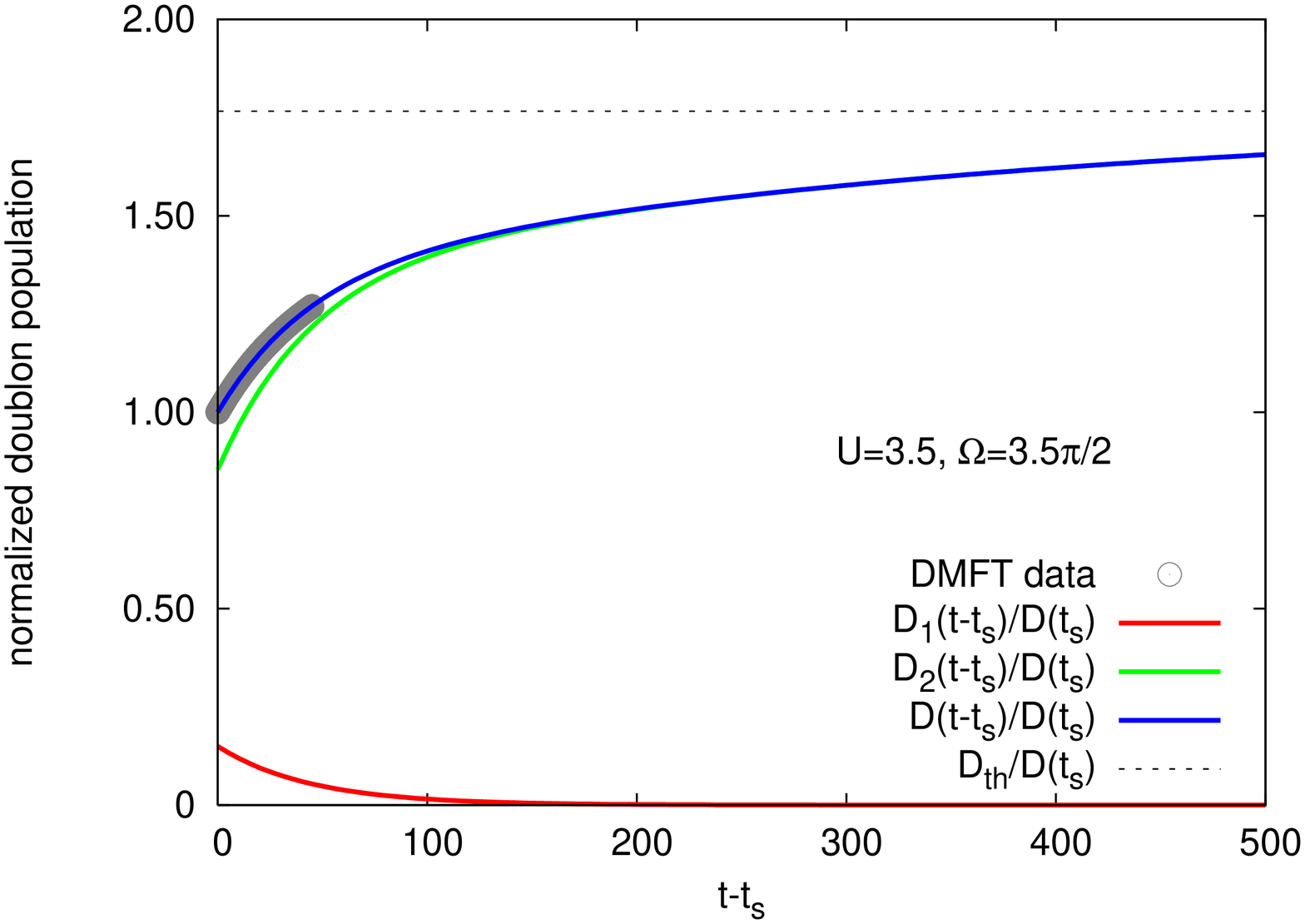}
\caption{Top panel: doublon production rate plotted as a function of $D_\text{th}-D(t)$ for different pulse energies and indicated values of $U$. Fits of model (\ref{model}) to the curves corresponding to the highest pulse energy are shown by the dashed lines. Bottom panel: Time evolution of the normalized doublon population as predicted from the fit to model (\ref{model}) for $U=3.5$, $\Omega=3.5\frac{\pi}{2}$ ($t_s=15$).
}
\label{doubling_u4}
\end{center}
\end{figure}

An instructive way to illustrate the two-step relaxation is to plot the doublon production rate $(d/dt)D(t)$ as a function of the deviation of the doublon density from the thermal value, $D_\text{th}-D(t)$. In this case, our model predicts a crossover from a small linear slope (corresponding to the slow long-time thermalization process) to a steeper slope (corresponding to the impact ionization processes). Indeed, for $U\gtrsim 3$, the data sets for different pulse energies fall roughly onto a single curve which describes such a crossover (upper panel of Fig.~\ref{doubling_u4}). For the data sets corresponding to the highest pulse frequencies, we plot the fits to model (\ref{model}) by dashed lines. These fits also roughly reproduce the relaxation for the other pulse frequencies, which shows that the model provides a consistent description of the thermalization process. 

In the lower panel of Fig.~\ref{doubling_u4}, we show the time-evolution for $U=3.5$, $\Omega=\frac{3\pi}{2}$, as predicted by the model (parameters from Tab.~\ref{table}). One can clearly see the two-step relaxation to the thermal value (dashed line), with a rapid initial increase of the doublon density, linked to impact ionization, followed by a much slower thermalization. 
Even though the relative high energy population is small (about $15\%$ at $t_s=15$), the impact ionization process contributes about half of the additional doublons needed for thermalization.

\subsubsection{Fluence dependence}
\label{subsec-fluence}

The impact ionization processes can be distinguished from the slower thermalization processes also by analyzing the dependence of the relaxation times on the photo-doping concentration, or fluence. Since impact ionization 
involves only a single doublon or hole in the initial state,  
we expect a weak fluence dependence of the fast relaxation time $\gamma$. On the other hand, the higher-order scattering processes that 
increase the number of doublons involve several doublons and/or holes. Hence these processes should exhibit a stronger dependence on the photo-doping concentration, so that we expect an increase in the slow relaxation time $\tau$ as the pulse amplitude is decreased.

We analyze the fluence dependence of the relaxation for $U=3$, $\Omega=3.5\pi/2$ and pulse amplitudes ranging from 0.25 to 6. The doping concentrations in the thermalized state and at $t=15$, shortly after the pulse, are given in Tab.~\ref{table_doping}. For small pulse amplitude, the number of photo-doped carriers grows proportional to the square of the pulse amplitude, as expected. The thermalization in this regime leads to more than a doubling of the mobile carriers. For pulse amplitudes $\gtrsim 2$, the number of carriers grows more slowly than the power of the field pulse, and also the relative increase of the doublon population associated with thermalization is lower. To avoid complications due to strongly non-linear absorption processes, we do not consider higher amplitudes.

\begin{figure}[t]
\begin{center}
\includegraphics[angle=-90, width=\columnwidth]{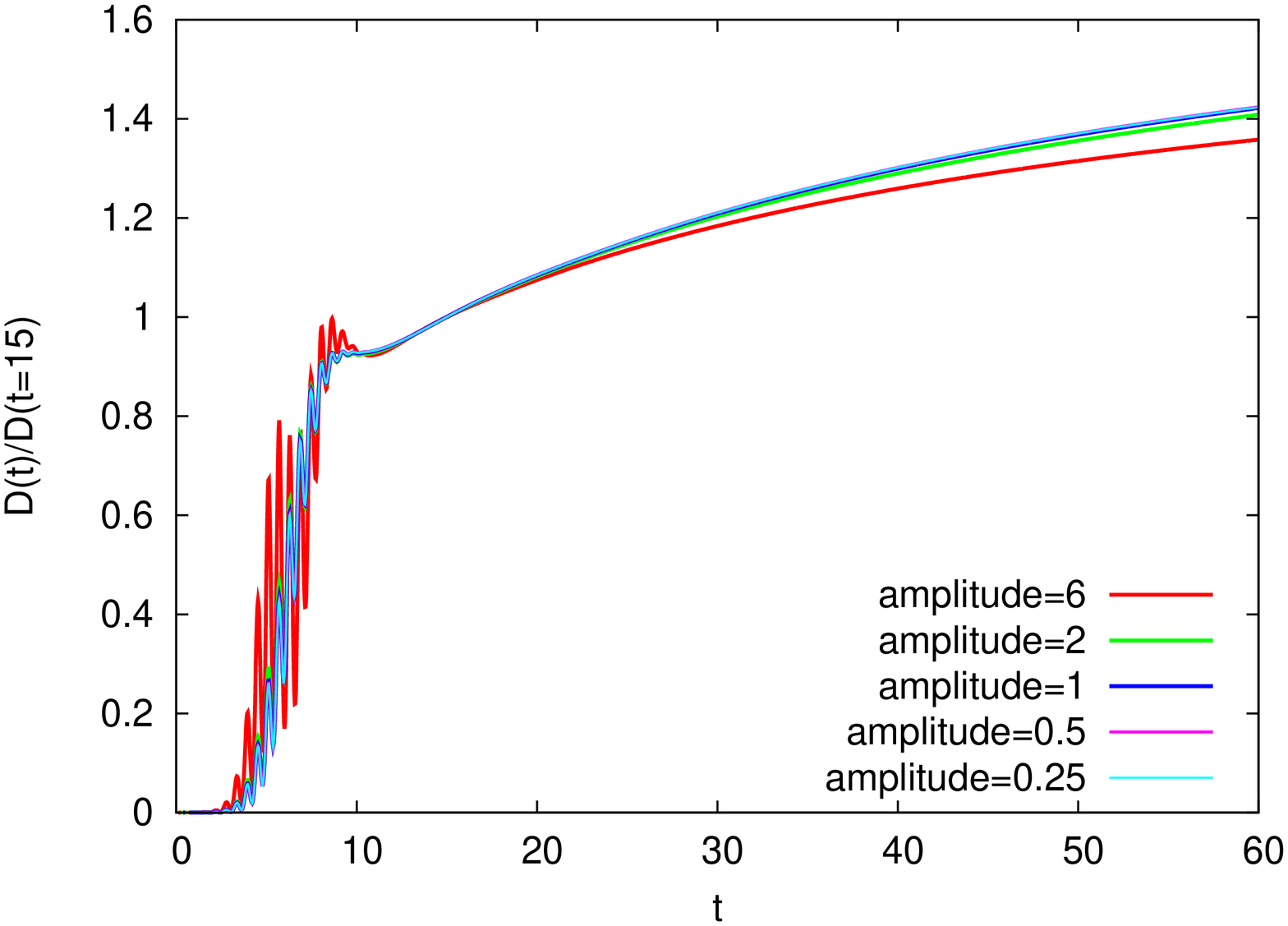}
\includegraphics[angle=-90, width=\columnwidth]{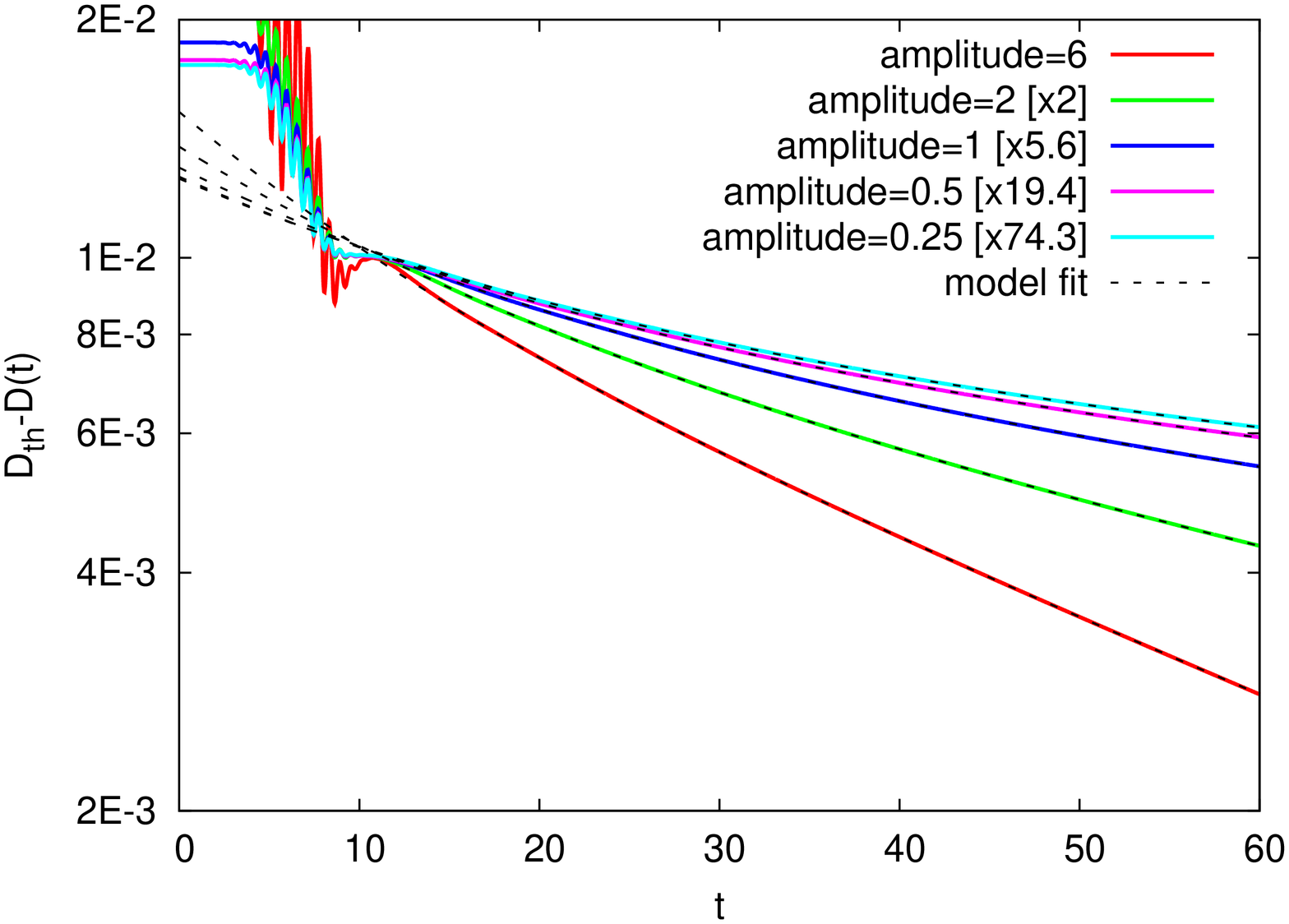}
\caption{Time evolution of the doublon concentration for $U=3$, $\Omega=3.5\pi/2$ and different pulse amplitudes. 
Top panel: Normalized doublon population. 
Bottom panel: Relaxation of the doublon concentration to the thermal value. 
The curves for amplitude $<6$ are multiplied by an arbitrary factor, 
to enable a better comparison of the long-time behavior. 
Dashed lines are fits to model (\ref{model}) on the time-interval $t\in[15,60]$. 
}
\label{twostep_doping}
\end{center}
\end{figure}

\begin{table}[b]
\begin{center}
\begin{tabular}{llllll}
amplitude\hspace{1mm}\mbox{} & $D(t_s)$\hspace{9mm}\mbox{}  & $D_\text{th}$\hspace{11mm}\mbox{} & $\beta_\text{th}$\hspace{7mm}\mbox{} & $\gamma$\hspace{7mm}\mbox{} & $\tau$\hspace{7mm}\mbox{}  \\ 
\hline
0.25 	& 0.000108	& $0.000236$ 	& 4.884 &		19.9 	& 	214	 \\ 
0.5 	& 0.000429	& $0.000917$ 	& 4.593 &		19.5	& 	194	 \\ 
1 	& 0.00167	& $0.00334$ 	& 3.879 &		18.3	& 	147 	 \\ 
2 	& 0.00593	& $0.0105$ 		& 2.854 & 		15.6 	& 	85.0  \\ 
6	& 0.0165 & $0.0252$ 		& 1.996 &		11.2 & 	46.1  \\ 
\hline
\end{tabular}
\end{center}
\caption{Relaxation times and initial populations extracted from fits to model (\ref{model}) in the range $t\in [15,60]$ for $U=3$, $\Omega=3.5\pi/2$, $t_s=15$ and indicated pulse amplitudes. 
}
\label{table_doping}
\end{table}

The top panel of Fig.~\ref{twostep_doping} shows the corresponding time evolution of the doublon concentration, normalized at $t=15$.  
The results for amplitudes smaller than 2 all collapse onto a single curve. This shows that in the initial stage of the relaxation, the doublon-hole production becomes independent of the doping concentration - a result consistent with a time evolution which is dominated by impact ionization. 
To see that the slow timescale is indeed more strongly dependent on fluence, we plot in the lower panel the difference to the thermal value, $D_\text{th}-D(t)$ on a logarithmic scale. 
To extract the two relaxation times $\gamma$ and $\tau$, we performed fits of the DMFT data with model (\ref{model}) and $t_s=15$. The slow timescale $\tau$ increases from about 50 to about 200 as the pulse amplitude is lowered from 6 to 0.25, while the fast timescale increases from about 10 to about 20 
(Tab.~\ref{table_doping}). 
For pulse amplitudes smaller than 2, i.e. in the small doping regime, the fast timescale becomes essentially independent of the doping concentration, while the slow timescale shows no sign of saturation and continues to increase with decreasing doping concentration. (Also the relative high-energy population, $D_1(t_s)/D(t_s)$, increases.) 
The estimated value of $\gamma\approx 20 \ll \tau$ implies that the initial fast increase of the doublon population evident in the upper panel of  Fig.~\ref{twostep_doping} is due to impact ionization. 

\begin{figure}[t]
\begin{center}
\includegraphics[angle=-90, width=\columnwidth]{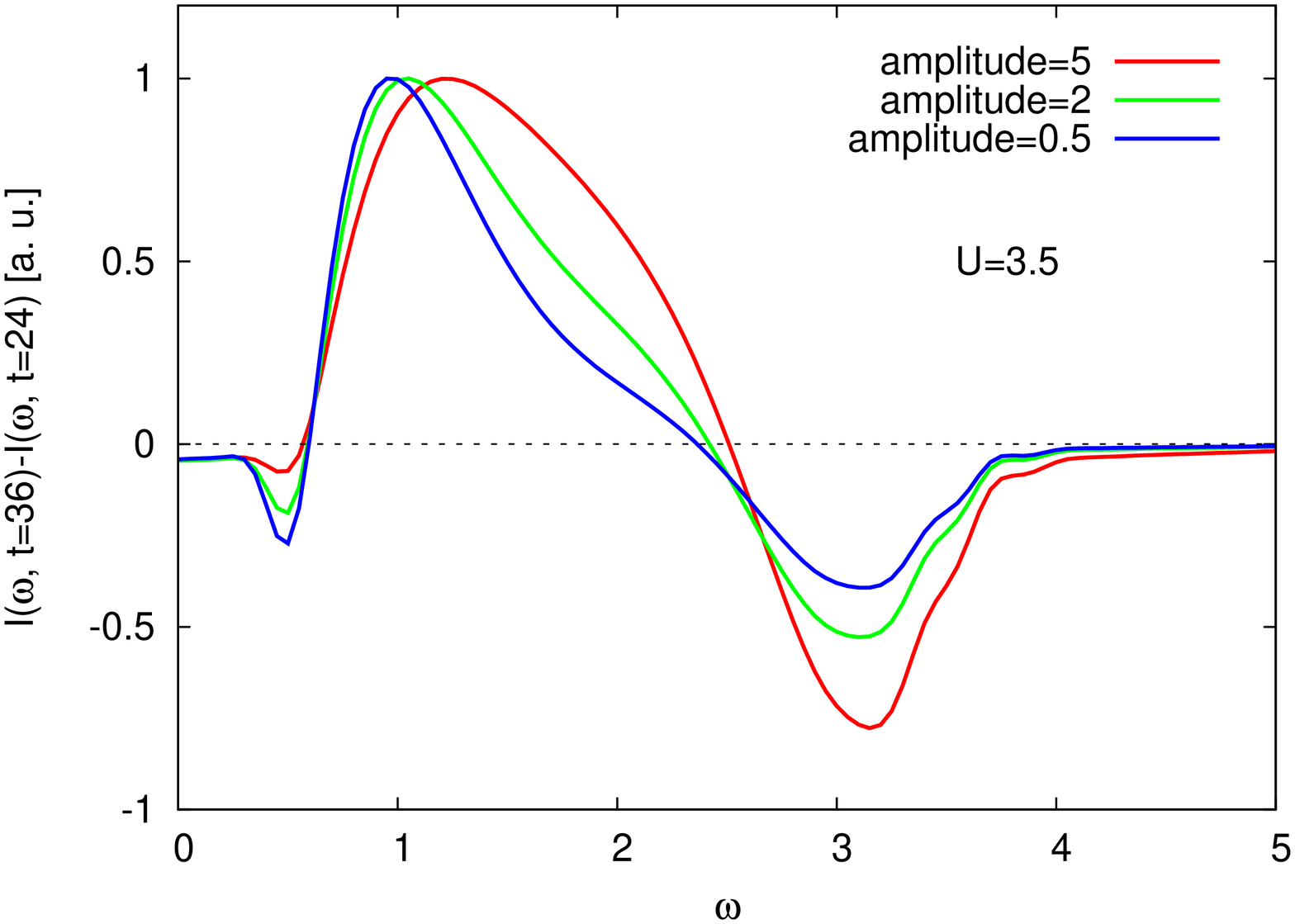}
\includegraphics[angle=-90, width=\columnwidth]{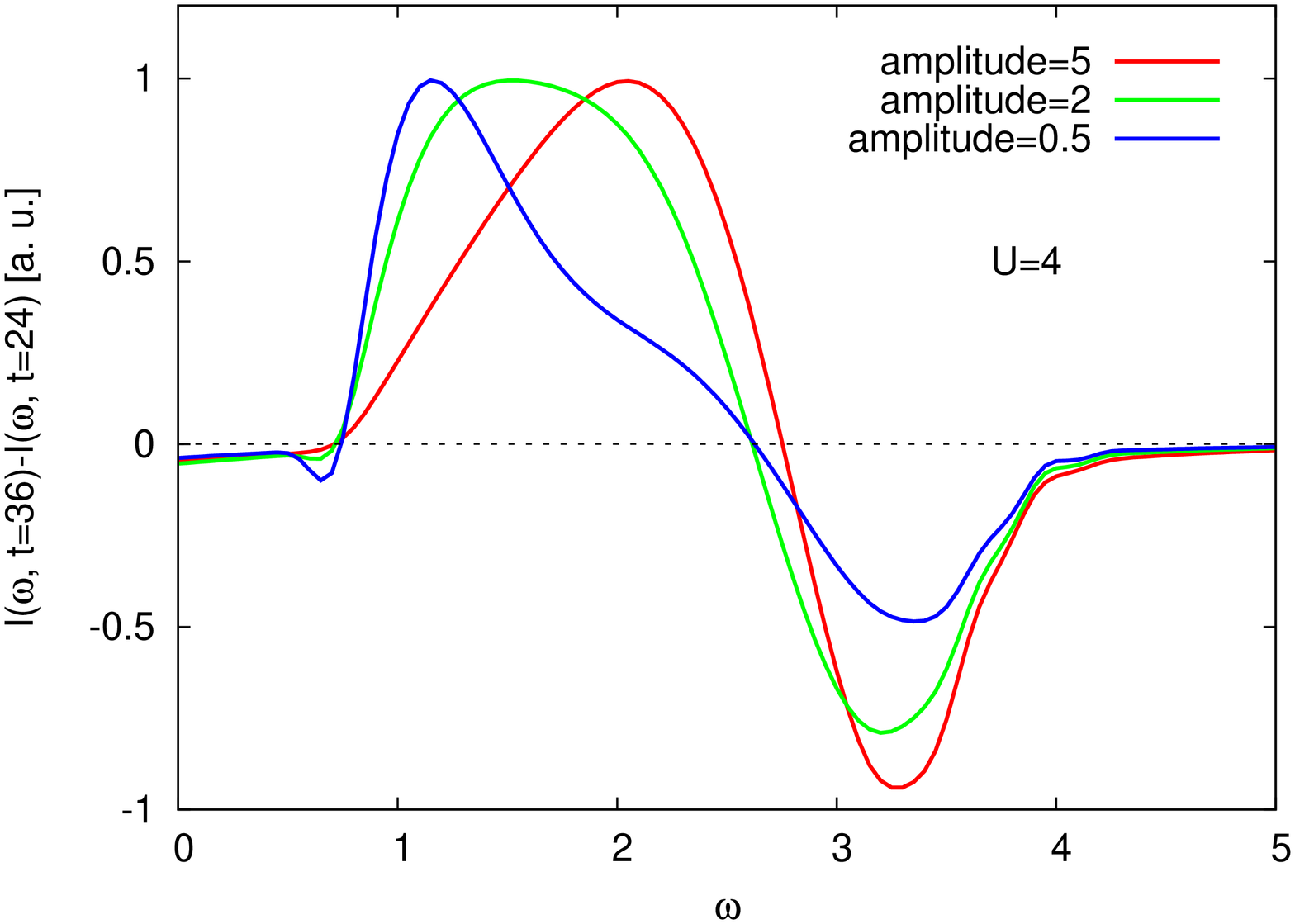}
\caption{Difference between the time resolved photoemission spectra measured at $t=36$ and $t=24$ for $\Omega=4\pi/2$ and the indicated values of the pulse amplitude. Top panel: $U=3.5$. Bottom panel: $U=4$.  For better comparison, the curves have been normalized such that the maximum difference is $1$. For $U=3.5$, the doublon concentration at $t=12$ is $D=0.0065$, 0.0021 and 0.00015 for amplitude $5$, $2$ and $0.5$, respectively. For $U=4$, the corresponding numbers are $D=0.014$, $0.0043$ and $0.00030$. 
}
\label{pes_amplitude}
\end{center}
\end{figure}

One also finds a fluence dependence in the time-resolved photoemission spectra. If the bandwidth, gap size and pump-pulse energy are compatible with impact ionization, then these processes dominate the doublon production and redistribution of spectral weight if the density of carriers is small.  As discussed in Sec.~\ref{subsec_pulse_freq}, a characteristic signature in the photoemission spectrum is an increase in spectral weight at  an energy $\omega_\text{gain}$ near the lower band egde, and a simultaneous decrease at an energy $\omega_\text{loss}$, 
where about three low-energy doublons are produced per high-energy doublon. 
This behavior is clearly evident for pulses with small amplitude in Fig.~\ref{pes_amplitude}. The top panel shows results for $U=3.5$ and $\Omega=4\pi/2$. For the blue curve ($D(t=12)=0.00015$), the area under the positive hump near the lower band edge is $2.5$ times larger than the area under negative hump at high energies. For the green curve ($D(t=12)=0.0021$) it is $2.3$ time larger. This means that for each doublon which disappears at high energy, more than two doublons are created at low energy, and thus we conclude that most of the doublons appearing near $\omega_\text{gain}$ are produced by impact ionization. 
A similar result is also displayed in the lower panel ($U=4$, $\Omega=4\pi/2$), where for the smallest pulse amplitude ($D(t=12)=0.0003$) we see two well-defined peaks with $\omega_\text{gain}/\omega_\text{loss}\approx 3$. 

With increasing fluence, the low energy hump broadens, which means that doublons appear at energies in the middle of the band, which are no longer compatible with impact ionization. They are instead the result of scattering processes between high-energy and low-energy doublons (or holes) which conserve the number of carriers. The red curve in the upper panel of Fig.~\ref{pes_amplitude}, which corresponds to $D(t=12)=0.0065$, and the green curve in the lower panel ($D(t=12)=0.0043$) show that these processes become relevant already at a doublon density of $\sim 0.5\%$. In these simulations, the area under the low-energy hump is about a factor of 2 larger than the area under the negative hump, which indicates that for each impact ionization process $\text{doublon}_\text{high} \rightarrow 3 \; \text{doublon}_\text{low}$ we have approximately also one scattering between a high energy and a low energy doublon (or hole), $\text{doublon}_\text{high} + \text{doublon}_\text{low}\rightarrow \; 2 \; \text{doublon}_\text{intermediate}$. Because the scattering probability of the latter process  is proportional to the carrier concentration, it can dominate the 
redistribution of spectral weight at even larger fluence (see red curve in the bottom panel).  

We note that deviations from the universal low-fluence evolution of the relative doublon concentration also appeared for doping concentrations larger than $\sim 0.5\%$ (see Fig.~\ref{twostep_doping} and Tab.~\ref{table_doping}). Thus our analysis of the photoemission spectra supports the interpretation that the universal curve is entirely controlled by impact ionization, while the slower increase of the relative doublon population seen for larger fluence is the result of competing scattering processes which deplete the high energy population.

\section{Discussion}
\label{sec-discussion}

In this study, we considered the thermalization dynamics after a photo-doping pulse in a purely electronic model without magnetic order.  
For a photo-excited Mott insulator, it is instructive to distinguish relaxation
processes (i) within the Hubbard band (which keep the number of
doublon-hole pairs fixed) and (ii) relaxation processes across the Mott
gap (which change the number of doublon-hole pairs). The simplest
relaxation process of type  (i) is electron-electron scattering, which
here is a  doublon-doublon, hole-hole or doublon-hole 
scattering. It keeps the number of doublons and holes fixed but transfers
energy from one doublon (hole) to another.  This process requires a
second doublon (or hole), and hence the corresponding relaxation rate of a carrier will
be proportional to the number of doublons (holes). Such processes redestribute the
energy among the doublons (holes) but do not change the doublon number. 

In order to change the number of doublons, 
type (ii) processes are needed. Again the simplest process is
electron-electron scattering which in this case corresponds to impact
ionization: a doublon (or hole) excites an electron across the Mott gap,
creating an additional doublon-hole pair. 
From one doublon, we obtain two doublons and one hole. If we also consider
the symmetric process for holes, 
impact ionization leads to a three-fold increase in the number of doublons and holes. 
These processes do not involve
other doublons (or holes) but the doublon kinetic energy must exceed the
size of the Mott gap. If the doublon energy is not large enough, only
less likely multi-scattering events can thermalize the number of
doublons.

We have discussed the characteristic signatures of impact ionization
in situations where the pulse energy is large and the gap is small, 
 as in Fig.\ \ref{doubling}: In this situation 
the photo-excited high-energy doublons create additional doublon-hole
pairs so that the number of doublons almost triples,
and a second peak develops in the photoemission spectrum at 
an energy corresponding to about 1/3 of the photo-excited high-energy peak.
Our data analysis based on the model Eqs.\ (\ref{model1})-(\ref{model3}) assumed 
that the high-energy doublon population $D_1$ decays only via impact ionization and
gives a good fit in these cases. We also find that impact ionization is fast for a Mott insulator. 
For a bandwidth $W$ of the order of 1~eV, corresponding to a unit of time of 0.66~fs, the fast 
relaxation times in Tab.~\ref{table} are of the order of 5-70 fs.

We have also seen that processes of type (i) have the potential to prevent impact
ionization by lowering the doublon energy before impact ionization
occurs.   Doublon-doublon scattering processes  become more
important when the number of doublons
is large (Fig.\ \ref{pes_amplitude}). 
In this case the rate
Eqs.\ (\ref{model1})  and (\ref{model2}) should be extended to 
$(dD_1/dt)_\text{imp+scat}=-(1/\gamma) D_1 - (1/\eta) D_1 D_2$ and
$(dD_2/dt)_\text{imp+scat}=+(3/\gamma) D_1 + (1 /\eta) D_1 D_2$
where $\eta$ is the relaxation time for scattering processes of type (i). 

In a real material further relaxation processes not considered in
our paper are possibly important, in particular phonon and (para)magnon scattering. In many cases, 
Mott gaps are of the order of 1 eV so that phonons and magnons
have a lower energy and can hence 
only contribute to type (i) processes. But as discussed above, this has the potential to 
prevent impact ionization.
In a Mott insulator with strong electron-phonon coupling, the cooling rate associated with electron-phonon scattering can be of the same order of magnitude as impact ionization.\cite{Werner2014}
For most systems  electron-phonons relaxation occurs however only on the 0.1-1ps time scale, which means that
these processes are slower than the observed impact ionization
in a Mott insulator. This is completely opposite to the behavior in semiconductors, where impact ionization has a much larger time scale than electron-phonon scattering.\cite{SchockleyQueisser61,SCimpactps}  For semiconductors, electron-phonon coupling hence
prevents impact ionization altogether.  

Even in a purely electronic system, additional relaxation processes may come into play.
At low temperature, in the magnetically ordered phase, spin-flip scattering provides a particularly efficient dissipation channel, which can lead to a fast redistribution of spectral weight within the Hubbard bands. Exact diagonalization based studies of the motion of a single carrier in an antiferromagnetic background suggest that the excess kinetic energy of a photo-doped carrier is transferred to the spin background within a few hopping times,\cite{Golez2014} and recent DMFT studies of photodoped antiferromagnetic Mott insulators revealed a very fast cooling of the photo-carriers.\cite{Werner2012, Eckstein2014} 
Also in the vicinity of an antiferromagnetic phase, 
short-range spin correlations provide 
an efficient scattering mechanism,\cite{Eckstein2014b}
whereas in one dimension the energy transfer to the spin system
seems to be inefficient.\cite{Al-Hassanieh08} 

In a model which takes into account the absorption of excess doublon kinetic energy by
phonon or magnon scattering, we have 
a reduction of the high-energy population $D_1$ and a corresponding increase of the low-energy population $D_2$. However, this time
these processes do not depend on the number
of doublons, hence Eq.~(\ref{model1}) has to be modified as $(dD_1/dt)_\text{imp+ph/mag}=(-1/\gamma-1/\kappa) D_1$, and Eq.~(\ref{model2}) as $(dD_2/dt)_\text{imp+ph/mag}=(3/\gamma+1/\kappa) D_1$, where  $\kappa$ is the corresponding
relaxation time.
  The efficient dissipation of kinetic energy and the associated rapid decrease in the high-energy population in an antiferromagnetic system is expected to have a significant effect on the thermalization dynamics in small gap Mott insulators. It reduces the effectiveness of the impact ionization process, leads to a slower adjustment of the doublon population, and thus a slower electronic thermalization. In the present work, we have chosen however a high
temperature 
where spin correlations are reduced, so that 
impact ionization can be more clearly identified.    
Our results should also be relevant at higher fluence, independent of magnetic ordering, because in this case the photo-doping leads to a rapid melting of antiferromagnetic correlations.\cite{Eckstein2014}

\section{Conclusion and Outlook}
The main finding of this study is that in situations where the 
gap-size is smaller than the width of the Hubbard bands, the kinetic energy of the photo-doped particles can be large enough that impact ionization processes play an important role in the initial relaxation. In fact, for the largest interactions considered ($U=3.5$-$4$), the doublon-hole production on the computationally accessible timescales is almost entirely due to impact ionization processes. We have demonstrated this by analyzing the time-resolved photoemission spectrum, and by extracting the impact-ionization and thermalization timescales from fits to a model with two exponentials, which was found to provide a rather good description of  the time evolution of the doublon density. 
These timescales depend on the gap size, with the slow timescale (related to higher order scattering processes) growing much more rapidly with gap size than the fast one (related to impact ionization), while the pulse frequency mainly affects the relative population of high-energy carriers which can trigger impact ionizations. The two timescales also exhibit a different dependence on the pulse amplitude (or density of photo-doped carriers): impact ionization process are insensitive to the doping concentration in the small-doping regime, while the slow timescale grows rapidly with decreasing fluence. For higher photo-doping, impact ionization can be masked and suppressed by doublon-doublon scattering.

Impact ionization may be relevant for Mott solar cell applications.
In the case of conventional semiconductor solar cells, the Coulomb 
interaction is weak so that 
interaction scattering (impact ionization) 
can hardly excite an electron across the 
semiconducting gap, i.e., create an additional electron-hole pair.
Impact ionization only occurs
on time scales of 1-100 ps, \cite{SCimpactps}
which is much longer than the typical time scales of 0.1-1 ps 
for electron-phonon scattering. Hence, for a conventional semiconductor, almost 
all the excess kinetic energy of photoinduced carriers
is transferred to lattice vibrations (heat). Consequently, for
each photo-excited electron-hole pair only the gap size
is harvested as an electrical energy quantum, independent of the energy quantum of the photon. This severely restricts the 
efficiency of semiconductor solar cells to about $31\%$, known
as the Schockley-Queisser limit.\cite{SchockleyQueisser61}
To overcome this limit,
solar cell quantum dots, e.g., based on
 PbSe nanocrystals, where larger Coulomb interactions and phonon-bottlenecks
effects can enhance impact ionization have been proposed.\cite{Schaller2004}
Also for Mott insulators the possibility of impact ionization has been 
discussed based on Fermi's Golden rule calculations for the Hubbard model.\cite{Manousakis2010}

In our paper we have shown that impact ionization 
in a Mott insulator can occur on time scales of the order of 
10 fs, i.e., much faster than typical electron-phonon relaxation times.
 Impact ionization processes are efficient only in Mott insulators with a small gap relative to the width of the Hubbard bands. This is quite difficult to realize for a one-band Hubbard model. However in multi-band Hubbard models or in charge-transfer insulators, the size of the gap can be much smaller than the width of the Hubbard and charge transfer bands, respectively.
Whether or not impact ionization can contribute significantly to the power produced by Mott solar cells such as the  recently proposed   LaVO$_3$-based heterostructure\cite{Assmann2013} remains an open question. To address this issue one would have to consider a realistic set-up, and also study the diffusion of the photo-doped carriers to the leads,\cite{Eckstein2013inhom} the effect of the spin background,\cite{Eckstein2014} and the coupling to phonons.\cite{Werner2013}     
In any case, our study has shown that  impact ionization in Mott insulators can be fast and can contribute effectively to the production of carriers. Hence,  Mott insulators have a potential 
to overcome the Schockley-Queisser limit, by harvesting more than the gap energy per photon. This class of materials can thus be expected to play an important role in the future development of highly efficient solar cells.

\vspace{5mm}

\acknowledgements
We thank T. Oka for stimulating discussions. The calculations were run on the UniFr cluster. PW is supported by FP7/ERC starting grant No. 278023, KH by the Austrian Science Fund SFB ViCoM F41.

\end{document}